\documentclass[12pt,twoside,fleqn]{article}

\usepackage{a4wide,cite}
\usepackage{amsmath,amssymb} 
\usepackage{graphicx}
\usepackage{color}

\numberwithin{equation}{section}
\allowdisplaybreaks[4]

\newcommand{\be}{\begin{equation}}
\newcommand{\ee}{\end{equation}}
\newcommand{\bea}{\begin{eqnarray}}
\newcommand{\eea}{\end{eqnarray}}

\def \nn{\nonumber}
\newcommand{\MS}{\ensuremath{\overline{\text{MS}}}}

\begin{document}

\begin{titlepage}

\centerline{\Large\bf NLL QCD contribution of the Electromagnetic}
\vspace{2mm}
\centerline{\boldmath \Large\bf  Dipole operator to
$\bar{B}\to X_s\gamma \gamma$ }
\vspace{2.5cm}

\begin{center}
  {\bf H.M.~Asatrian$^a$, C.~Greub$^b$, A.~Kokulu$^b$, A.~Yeghiazaryan$^a$}\\[1mm]
  {$^a$\sl Yerevan Physics Institute, 0036 Yerevan, Armenia}\\[1mm]
  {$^b$\sl Albert Einstein Center for Fundamental Physics, Institute for Theoretical Physics,\\
    Univ.~of Bern, CH-3012 Bern, Switzerland}\\[1mm]
 
\end{center}

\medskip

\vspace{2cm}
\begin{abstract}
\noindent 
We calculate the set of $O(\alpha_s)$ corrections to the double 
differential decay width \\  
$d\Gamma_{77}/(ds_1 \, ds_2)$
for the process $\bar{B} \to X_s \gamma \gamma$ originating from
diagrams involving the electromagnetic dipole operator
${\cal O}_7$. The kinematical variables $s_1$ and $s_2$ are defined as
$s_i=(p_b - q_i)^2/m_b^2$, where $p_b$, $q_1$, $q_2$ are the momenta
of $b$-quark and two photons. While the (renormalized) virtual
corrections are worked exactly for a certain range of $s_1$ and $s_2$, 
we retain in the gluon bremsstrahlung process only the leading power w.r.t. the (normalized)
hadronic mass $s_3=(p_b-q_1-q_2)^2/m_b^2$ in the underlying triple
differential decay width $d\Gamma_{77}/(ds_1 ds_2 ds_3)$. The double
differential decay width, based on this approximation is free of
infrared- and collinear singularities when combining virtual- and
bremsstrahlung corrections. The corresponding results are obtained 
analytically. When retaining all powers in $s_3$, the sum of virtual-
and bremstrahlung corrections contains uncanceled $1/\epsilon$
singularities (which are due to collinear {\it photon} emission from the
$s$-quark) and other concepts, which go beyond
perturbation theory, like parton fragmentation functions of a quark or
a gluon into a photon, are needed which is beyond the scope of our paper.

\end{abstract}

\end{titlepage}

\section{Introduction}
Inclusive rare $B$-meson decays are known to be a unique source of indirect information about 
physics at scales of several hundred GeV. In the Standard Model (SM) all these processes 
proceed through loop diagrams and thus are relatively suppressed. In the extensions 
of the SM the contributions stemming from the diagrams with ``new'' 
particles in the loops can be comparable or even larger than the contribution from 
the SM. Thus getting experimental information on rare decays puts strong 
constraints on the extensions of the SM or can even lead to a  
disagreement with the SM predictions, providing evidence for some ``new physics''. 

To make a rigorous comparison between experiment and theory, precise
SM calculations for the (differential) decay rates are mandatory. While the
branching ratios for $\bar{B} \to X_s \gamma$ \cite{Misiak:2006zs}
and $\bar{B} \to X_s \ell^+
\ell^-$ are known today even to
next-to-next-to-leading logarithmic (NNLL) precision (for reviews, see
\cite{Hurth:2010tk,Buras:2011we}),
other branching ratios, like the one for $\bar{B} \to X_s \gamma
\gamma$ discussed in this paper, are only known to leading logarithmic
(LL) precision in the SM
\cite{Simma:1990nr,Reina:1996up,Reina:1997my,Cao:2001uj}.
In contrast to $\bar{B} \to X_s
\gamma$, the current-current operator ${\cal O}_2$ has a non-vanishing matrix
element for $b \to s \gamma \gamma$ at order $\alpha_s^0$ precision, 
leading to an interesting interference pattern with the contributions associated
with the electromagnetic dipole operator ${\cal O}_7$ already at LL
precision. As a consequence, potential new physics should be clearly visible
not only in the total branching ratio, but also in the 
differential distributions.
 
As the process $\bar{B} \to X_s \gamma \gamma$ is expected to be measured at
the planned Super $B$-factories in Japan and Italy, it is necessary
to calculate the differential distributions to NLL precision in the
SM, in order to
fully exploit its potential concerning new physics. 
The starting point of our calculation is the effective Hamiltonian,
obtained by integrating out the heavy particles in the SM, leading to
\be
 {\cal H}_{eff} = - \frac{4 G_F}{\sqrt{2}} \,V_{ts}^\star V_{tb} 
   \sum_{i=1}^8 C_i(\mu) {\cal O}_i(\mu)  \, ,
\label{Heff}
\ee
where we use the operator basis introduced in \cite{Chetyrkin:1996vx}:
\be
\begin{array}{llll}
{\cal O}_1 \,= &\!
 (\bar{s}_L \gamma_\mu T^a c_L)\, 
 (\bar{c}_L \gamma^\mu T_a b_L)\,, 
               &  \quad 
{\cal O}_2 \,= &\!
 (\bar{s}_L \gamma_\mu c_L)\, 
 (\bar{c}_L \gamma^\mu b_L)\,,   \\[1.002ex]
{\cal O}_3 \,= &\!
 (\bar{s}_L \gamma_\mu b_L) 
 \sum_q
 (\bar{q} \gamma^\mu q)\,, 
               &  \quad 
{\cal O}_4 \,= &\!
 (\bar{s}_L \gamma_\mu T^a b_L) 
 \sum_q
 (\bar{q} \gamma^\mu T_a q)\,,  \\[1.002ex]
{\cal O}_5 \,= &\!
 (\bar{s}_L \gamma_\mu \gamma_\nu \gamma_\rho b_L) 
 \sum_q
 (\bar{q} \gamma^\mu \gamma^\nu \gamma^\rho q)\,, 
               &  \quad 
{\cal O}_6 \,= &\!
 (\bar{s}_L \gamma_\mu \gamma_\nu \gamma_\rho T^a b_L) 
 \sum_q
 (\bar{q} \gamma^\mu \gamma^\nu \gamma^\rho T_a q)\,,  \\[1.002ex]
{\cal O}_7 \,= &\!
  \frac{e}{16\pi^2} \,\bar{m}_b(\mu) \,
 (\bar{s}_L \sigma^{\mu\nu} b_R) \, F_{\mu\nu}\,, 
               &  \quad 
{\cal O}_8 \,= &\!
  \frac{g_s}{16\pi^2} \,\bar{m}_b(\mu) \,
 (\bar{s}_L \sigma^{\mu\nu} T^a b_R)
     \, G^a_{\mu\nu}\, .
\end{array} 
\label{opbasis}
\ee
The symbols $T^a$ ($a=1,8$) denote the $SU(3)$ color generators; 
$g_s$ and $e$, the strong and electromagnetic coupling constants.
In eq.~(\ref{opbasis}), 
$\bar{m}_b(\mu)$ is the running $b$-quark mass 
in the $\MS$-scheme at the renormalization scale $\mu$.
As we are not interested in CP-violation effects in the present paper, we 
made use of the approximation
 $V_{ub} V_{us}^* \ll V_{tb} V_{ts}^* $ when writing
eq. (\ref{Heff}). Furthermore, we also put $m_s=0$.  

While the Wilson coefficients $C_i(\mu)$ appearing in eq. (\ref{Heff})
are known to sufficient precision at the low scale $\mu \sim m_b$
since a long time (see e.g. the reviews \cite{Hurth:2010tk,Buras:2011we}
and references therein), the matrix elements 
$\langle s \gamma \gamma|{\cal  O}_i|b\rangle$ and 
$\langle s \gamma \gamma \, g|{\cal  O}_i|b\rangle$, 
which in a NLL calculation are needed to order
$g_s^2$ and $g_s$, respectively, are not known yet. To calculate the
$({\cal O}_i,{\cal O}_j)$-interference contributions for the
differential distributions at order
$\alpha_s$ is in many respects of similar complexity as the
calculation of the photon energy spectrum at order $\alpha_s^2$
needed for the NNLL computation. There, the individual
interference contributions, which all involve extensive calculations, were
published in separate papers, sometimes even by two independent groups
(see e.g. \cite{Melnikov:2005bx} and \cite{Asatrian:2006sm}).
It therefore cannot be expected that the NLL results for the
differential distributions related to $\bar{B} \to X_s \gamma \gamma$ are 
given in a single paper. As a first step in this NLL enterprise, we
derive in the present paper  the $O(\alpha_s)$
corrections 
to the $({\cal O}_7,{\cal O}_7)$-interference contribution to the double 
differential decay width $d\Gamma/(ds_1 ds_2)$ at the partonic level. 
The variables $s_1$
and $s_2$ are defined as $s_i=(p_b-q_i)^2/m_b^2$, where $p_b$
and $q_i$ denote the four-momenta of the $b$-quark and the two
photons, 
respectively.

At order $\alpha_s$
there are contributions to $d\Gamma_{77}/(ds_1 ds_2)$ with three
particles 
($s$-quark and two photons)
and four particles ($s$-quark, two photons and a gluon) in the final state.
These contributions correspond to specific cuts of the $b$-quarks
self-energy at order $\alpha^2 \times \alpha_s$, involving twice the
operator ${\cal O}_7$. As there are additional cuts, which contain for
example only one photon, our observable cannot be obtained using the
optical theorem, i.e., by taking the absorptive part of the $b$-quark
self-energy at three-loop. We therefore calculate the mentioned 
contributions with three and four particles in the final state individually.

As discussed in section \ref{sec:leadingorder}, we work out the QCD corrections
to the double differential decay width 
in the kinematical range 
\[
0 < s_1 < 1 \quad ; \quad 0 < s_2 < 1-s_1 \, .
\]
Concerning the virtual corrections, all singularities (after
ultra-violet renormalization) are due to  {\bf soft gluon} exchanges
and/or  {\bf collinear gluon} exchanges involving the $s$-quark. Concerning the
bremsstrahlung corrections (restricted to the same range of $s_1$ and
$s_2$), there are in addition kinematical situations where {\bf collinear photons}
are emitted from the $s$-quark. The corresponding singularities are not
canceled when combined with the virtual corrections, as discussed in
detail in section \ref{sec:bremsstrahlung}. We found, however, that there are no
singularities associated with collinear photon emission in the double
differential decay width when only retaining
the leading power w.r.t to the (normalized) hadronic mass
$s_3=(p_b - q_1 - q_2)^2/m_b^2$ in the underlying triple differential distribution
$d\Gamma_{77}/(ds_1 ds_2 ds_3)$. Our results of this paper are
obtained within this ``approximation''. When going beyond, other
concepts which go beyond perturbation theory, like parton
fragmentation functions of a quark or a gluon into a photon, are
needed \cite{Kapustin:1995fk}. 

Before moving to the detailed organization of our paper, we should
mention that the inclusive double radiative process $\bar{B} \to X_s \gamma
\gamma$ has also been explored in several extensions of the SM
\cite{Reina:1996up,Gemintern:2004bw,Cao:2001uj}. Also 
the corresponding exclusive modes, $B_s \to \gamma \gamma$  
and $B\to K \gamma \gamma$, have been 
examined before, both in the SM
\cite{Reina:1997my,Chang:1997fs,Hiller:1997ie,Bosch:2002bv,Bosch:2002bw,
Hiller:2004wc,Hiller:2005ga,Lin:1989vj,Herrlich:1991bq,Choudhury:2002yu} and in its extensions  \cite{Aliev:1997uz,Hiller:2004wc,Hiller:2005ga,Bertolini:1998hp,Gemintern:2004bw,Bigi:2006vc,Devidze:1998hy,Aliev:1993ea,Xiao:2003jn,XiuMei:2011iv,Huo:2003cj,Chen:2011te}.
We should add that the long-distance resonant effects were
also discussed in the literature (see e.g. \cite{Reina:1997my} and the
references therein). 
Finally, the effects of photon emission from the spectator quark in
the $B$-meson were discussed in
\cite{Chang:1997fs,Ignatiev:2003qm}. 

The remainder of this paper is organized as follows.
In section \ref{sec:leadingorder} we work out the double differential distribution 
$d\Gamma_{77}/(ds_1 ds_2)$ in leading order, i.e., without taking into
account QCD corrections to the matrix element 
$\langle s \gamma \gamma |{\cal  O}_7|b\rangle$. We retain, however,
terms up to order $\epsilon^1$,
with $\epsilon$ being the dimensional regulator ($d=4-2\epsilon$).
Section \ref{sec:virtual} is devoted to the calculation of the virtual  
corrections of order $\alpha_s$ to the double differential decay
width. In section \ref{sec:bremsstrahlung}
the corresponding gluon bremsstrahlung corrections to the double
differential width are worked out in
the approximation where only the leading power w.r.t. the (normalized)
hadronic mass $s_3$ is retained at the level of the triple
differential decay width $d\Gamma_{77}/(ds_1 ds_2 ds_3)$.  
In section \ref{sec:combination} virtual- and bremsstrahlung corrections are combined and
the result for the double differential decay width, which is free of
infrared- and collinear singularities, is given in
analytic form.  In section \ref{sec:numerics} we illustrate the numerical impact of the
NLL corrections and in section \ref{sec:technical} we present the technical details of
our calculations. The paper ends with a short summary in section \ref{sec:summary}.

\section{Leading order result}\label{sec:leadingorder}
\begin{figure}[h]
\begin{center}
\includegraphics[width=0.9\textwidth]{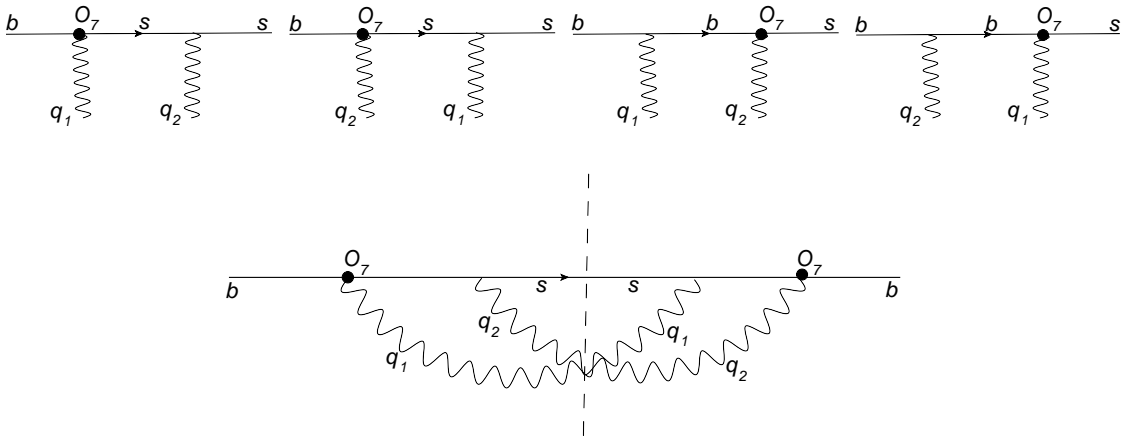}
\caption{On the first line the diagrams defining the tree-level
amplitude for $b \to s \gamma \gamma$ associated with ${\cal O}_7$ are
shown. The four-momenta of the $b$-quark, the $s$-quarks and the two
photons are denoted by $p_b$, $p_s$, $q_1$ and $q_2$, respectively.
On the second line the contribution to the decay width corresponding
to the interference of first and second diagram is shown.}
\label{fig:amplitudetree}
\end{center}
\end{figure}

\begin{figure}[h]
\begin{center}
\includegraphics[width=5.0cm]{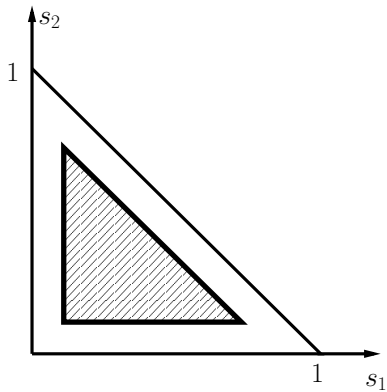}
\caption{The relevant phase-space region for $(s_1,s_2)$ used in this paper is 
shown by the shaded area.}
\label{fig:phasespace}
\end{center}
\end{figure}

In this section we discuss the double differential decay width
$d\Gamma_{77}/(ds_1 ds_2)$ at lowest order in QCD, i.e. $\alpha_s^0$.
The dimensionless variables $s_1$ and $s_2$ are defined everywhere in this
paper as
\be
s_1=\frac{(p_b-q_1)^2}{m_b^2} \quad ; \quad
s_2=\frac{(p_b-q_2)^2}{m_b^2} \, .
\label{kinematicvariables}
\ee
At lowest order the double differential decay width is based on the 
diagrams shown in Fig.~\ref{fig:amplitudetree}.
The variables
$s_1$ and $s_2$ form a complete set of kinematically independent
variables for the three-body decay $b \to s \gamma \gamma$.
Their kinematical range is as follows:
\[
0 \le s_1 \le 1 \quad ; \quad 0 \le s_2 \le 1-s_1 \, .
\]
The energies $E_1$ and $E_2$ in the rest-frame of the $b$-quark of the two photons are 
related to $s_1$ and $s_2$ in a simple way: $s_i=1-2 \, E_i/m_b$.
As the energies $E_i$ of the photons have to be away from zero in order to
be observed, the values of $s_1$ and $s_2$ can be considered to be
smaller than one. By additionally requiring $s_1$ and $s_2$ to be larger than zero,
we exclude collinear photon emission from the $s$-quark, because
$2 p_s q_1 = (p_s+q_1)^2=(p_b-q_2)^2=s_2 \, m_b^2>0$ and 
$2 p_s q_2=(p_s+q_2)^2=(p_b-q_1)^2=s_1 \, m_b^2>0$. It is also easy to
implement a lower cut on the invariant mass squared $s$ of the of the two 
photons by observing that $s=(q_1+q_2)^2=1-s_1-s_2$. To parametrize
all the mentioned conditions in terms of one parameter $c$ (with $c>0$),
one can proceed as suggested in \cite{Reina:1996up}:
\be
s_1 \ge c \, , \quad s_2 \ge c \, , \quad 1-s_1-s_2 \ge c \, .
\label{kinematical_cuts}
\ee   
Applying such cuts, the relevant phase-space region in the
$(s_1,s_2)$-plane is shown by the shaded area in
Fig.~\ref{fig:phasespace}. Our aim in this paper is to work out the
double differential decay width in this restricted area of the $s_1$ and
the $s_2$ variable also when discussing the gluon bremsstrahlung
corrections\footnote{In this case, the normalized invariant mass
squared $s$ of the two photons reads $s=1-s_1-s_2+s_3$, where
$s_3$ is the normalized hadronic mass squared. The condition
$1-s_1-s_2 \ge c$ then still eliminates two-photon configurations with
small invariant mass.}.
In this restricted region of the phase-space, the tree-level amplitude
is free of infrared- and collinear singularities. To exhibit the
singularity structure of the virtual corrections discussed in the
next section in a transparent way, it is useful to give the
leading-order spectrum
in $d=4-2\epsilon$ dimensions. We obtain
\begin{eqnarray}
&&\frac{d\Gamma_{77}^{(0,d)}}{ds_1 \, ds_2} = \frac{\alpha^2 \,
    \bar{m}_b^2(\mu) \, m_b^3 
\, |C_{7,eff}(\mu)|^2 \, G_F^2 \,
  |V_{tb} V_{ts}^*|^2 \,  Q_d^2}{1024 \, \pi^5} \, \left(
  \frac{\mu}{m_b}\right)^{4\epsilon}   r 
\label{treea}
\end{eqnarray}
with
\begin{eqnarray}
&& r=\frac{\left[r_0+\epsilon (r_1+r_2+r_3+r_4) \right]
   \left(1-s_1-s_2\right)}{\left(1-s_1\right)^2
   s_1 \left(1-s_2\right)^2 s_2} \, .
\label{treeb}  
\end{eqnarray} 
In $r$ we retained  terms of order $\epsilon^1$, while discarding terms of higher order.
The individual pieces $r_0,\ldots,r_4$ read  
\begin{eqnarray}
\nonumber r_0&=&-48 s_2^3 s_1^3+96 s_2^2 s_1^3-56 s_2 s_1^3+8
   s_1^3+96 s_2^3 s_1^2-192 s_2^2 s_1^2+112
   s_2 s_1^2-56 s_2^3 s_1+
   \\&&\nonumber 112 s_2^2 s_1-96 s_2
   s_1+8 s_1+8 s_2^3+8 s_2
    \end{eqnarray}
     \begin{eqnarray}
\nonumber r_1&=&-16 s_2^2 s_1^3+16 s_2 s_1^3-16 s_2^3 s_1^2+48
   s_2^2 s_1^2-32 s_2 s_1^2+16 s_1^2+16 s_2^3
   s_1- \\&&\nonumber 32 s_2^2 s_1-16 s_2 s_1+16 s_2^2
    \end{eqnarray}
   \begin{eqnarray}
\nonumber r_2&=&\left(48 s_2^3 s_1^3-96 s_2^2 s_1^3+56 s_2
   s_1^3-8 s_1^3-96 s_2^3 s_1^2+192 s_2^2
   s_1^2-
   \right.\\&& \left.\nonumber
   112 s_2 s_1^2+56 s_2^3 s_1-112 s_2^2
   s_1+96 s_2 s_1-8 s_1-8 s_2^3-8 s_2\right)
   \log \left(s_1\right)
    \end{eqnarray}
   \begin{eqnarray}
\nonumber r_3&=& r_2(s_1 \leftrightarrow s_2)
    \end{eqnarray}
    \begin{eqnarray}
\nonumber r_4&=&\left[48 s_2^3 s_1^3-96 s_2^2 s_1^3+56 s_2
   s_1^3-8 s_1^3-96 s_2^3 s_1^2+192 s_2^2
   s_1^2-112 s_2 s_1^2+56 s_2^3 s_1-
    \right.\\&& \left.\nonumber
   112 s_2^2 s_1+
   96 s_2 s_1-8 s_1-8 s_2^3-8 s_2\right]
   \log \left(1-s_1-s_2\right)
    \end{eqnarray}
In eq. (\ref{treea}) the symbols  $\bar{m}_b(\mu)$ and  $m_b$
denote the mass of the $b$-quark in the $\overline{\rm{MS}}$-scheme
and in the on-shell scheme, respectively.
 
In $d=4$ dimensions, the leading-order spectrum (in our restricted
phase-space) is obtained by simply
putting $\epsilon$ to zero, obtaining
\begin{eqnarray}
&&\frac{d\Gamma_{77}^{(0)}}{ds_1 \, ds_2} = \frac{\alpha^2 \, \bar{m}_b^2(\mu) \, m_b^3 \, |C_{7,eff}(\mu)|^2 \, G_F^2 \,
  |V_{tb} V_{ts}^*|^2 \,  Q_d^2}{1024 \, \pi^5} \, 
  \frac{(1-s_1-s_2)}{(1-s_1)^2 s_1 (1-s_2)^2 s_2} \, r_0 \, .
\label{treezero}
\end{eqnarray}
\section{Virtual corrections}\label{sec:virtual}
\begin{figure}[h]
\begin{center}
\includegraphics[width=0.9\textwidth]{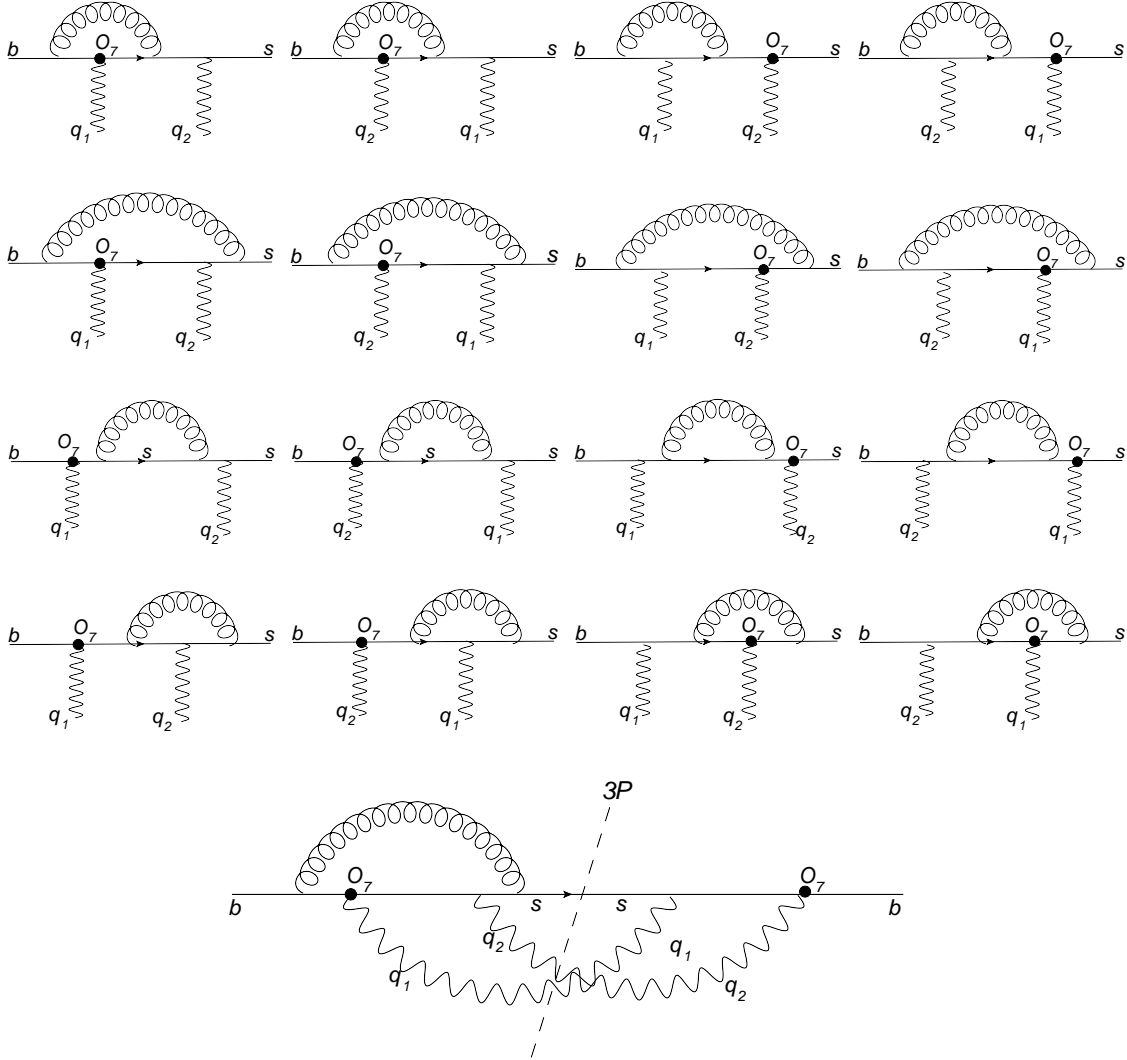}
\caption{On the first four lines the diagrams defining the one-loop
amplitude for $b \to s \gamma \gamma$ associated with ${\cal O}_7$ are
shown. Diagrams with self-energy insertions on the external quark-legs
are not shown.
On the last line the contribution to the decay width corresponding
to the interference of the first diagram on the second line with the second
(tree-level) diagram in Fig. \ref{fig:amplitudetree} is shown.}
\label{fig:amplitudevirtual}
\end{center}
\end{figure}

We now turn to the calculation of the virtual QCD corrections, i.e. to
the contributions of order $\alpha_s$ with three particles in the
final state. The diagrams defining the (unrenormalized) 
virtual corrections at the amplitude level are shown on the first four
lines of
Fig. \ref{fig:amplitudevirtual}. 
As the diagrams with a self-energy insertion on the external $b$- and
$s$-quark legs are taken into account in the renormalization process,
these diagrams are not shown in Fig. \ref{fig:amplitudevirtual}.
In order to get the (unrenormalized) virtual corrections 
$d\Gamma_{77}^{\rm bare}/(ds_1 ds_2)$ of order $\alpha_s$ to the
decay width, we have to work out the interference of the diagrams on the
first four lines in Fig. \ref{fig:amplitudevirtual} with the leading
order diagrams in Fig. \ref{fig:amplitudetree}. One of these
interference contributions is shown on the last line in Fig. \ref{fig:amplitudevirtual}. 
To illustrate the calculational procedure for getting the virtual
corrections to the decay width, we describe in section
\ref{subsec:virtual} the relevant steps for the
particular interference shown in Fig. \ref{fig:amplitudevirtual}. 

In addition, we have to work out the counterterm contributions 
to the decay width. They can be split into two parts, according to
\be
\frac{d\Gamma_{77}^{\rm ct}}{ds_1 ds_2}=
\frac{d\Gamma_{77}^{{\rm ct},(A)}}{ds_1 ds_2}+
\frac{d\Gamma_{77}^{{\rm ct},(B)}}{ds_1 ds_2} \, .
\ee
\newpage
\noindent
Part (A) involves the LSZ factors 
$\sqrt{Z_{2b}^{\rm OS}}$ and $\sqrt{Z_{2s}^{\rm OS}}$ for the $b$- and
$s$-quark field, as well as the
self-renormalization constant $Z_{77}^{\MS}$ of the operator ${\cal
  O}_7$ and $Z_{m_b}^{\MS}$ 
renormalizing the factor $\bar{m}_b(\mu)$ present in the
operator ${\cal O}_7$. The explicit results for these $Z$-factors are
given to relevant precision in Appendix \ref{append:renomalizationconstants}.
For part (A) we get
\be
\frac{d\Gamma_{77}^{{\rm ct},(A)}}{ds_1 ds_2} = 
\left[ \delta Z_{2b}^{\rm OS} + \delta Z_{2s}^{\rm OS} + 2 \, \delta Z_{m_b}^{\MS} +
  2 \, \delta Z_{77}^{\MS} \right] \, \frac{d\Gamma_{77}^{(0,d)}}{ds_1 ds_2} \, ,
\ee
where $d\Gamma_{77}^{(0,d)}/(ds_1 ds_2)$ is the leading order double
differential decay width in $d$-dimensions, as given in
eq. (\ref{treea}).

\begin{figure}[h]
\begin{center}
\includegraphics[width=0.6\textwidth]{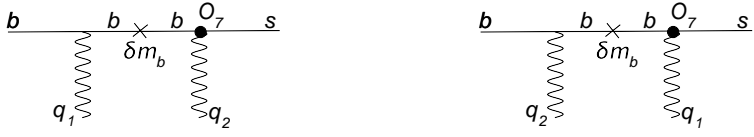}
\caption{Counterterm diagrams with a $\delta m_b$ insertion, see text.}
\label{fig:amplitudembinsertion}
\end{center}
\end{figure}
The counterterms defining part (B) are due to the insertion 
of $-i \delta m_b \bar b b$  
in the internal $b$-quark line in the leading order diagrams as
indicated in Fig. \ref{fig:amplitudembinsertion}, where 
\[
\delta m_b = (Z_{m_b}^{\rm OS}-1) \, m_b \, .
\]
More precisely, Part (B) consists of the interference of the diagrams in
Fig.~\ref{fig:amplitudembinsertion} with the leading order diagrams in 
Fig.~\ref{fig:amplitudetree}. 

By adding $d\Gamma_{77}^{\rm bare}/(ds_1 ds_2)$ and 
$d \Gamma_{77}^{{\rm ct}}/(ds_1 ds_2)$,
we get the result for the renormalized virtual corrections to the
spectrum, $d\Gamma_{77}^{(1),virt}/(ds_1 \, ds_2)$.
It is useful to decompose this result into two pieces,
\begin{equation}
\frac{d\Gamma_{77}^{(1),virt}}{ds_1 \, ds_2} = 
\frac{d\Gamma_{77}^{(1,a),virt}}{ds_1 \, ds_2} + 
\frac{d\Gamma_{77}^{(1,b),virt}}{ds_1 \, ds_2}  \, .
\end{equation}
The infrared- and collinear singularities are completely contained in
$d\Gamma_{77}^{(1,a),virt}/(ds_1 \, ds_2)$. Explicitly, we obtain
\begin{equation}
\frac{d\Gamma_{77}^{(1,a),virt}}{ds_1 \, ds_2} = \frac{\alpha_s}{4\pi}
\, C_F \,
\left(-\frac{2}{\epsilon^2} + \frac{4 \log (s_1+s_2)-5}{\epsilon}
\right) \, \left( \frac{\mu}{m_b}\right)^{2\epsilon} \, \frac{d
  \Gamma_{77}^{(0,d)}}{ds_1 \, ds_2} 
\label{virtuala} 
\end{equation}
where  $d\Gamma_{77}^{(0,d)}/(ds_1 \, ds_2)$ is understood to be taken
exactly as given in eqs. (\ref{treea}) and (\ref{treeb}), i.e., by including the terms of
order $\epsilon^1$ in $r$. From the explicit expression  
$d\Gamma_{77}^{(1,a),virt}/(ds_1 \, ds_2)$
we see that the singularity structure consists of a simple
singular factor multiplying the corresponding tree-level decay width
in $d$-dimensions. We stress that singularities (represented by
$1/\epsilon^2$ and $1/\epsilon$ poles) are entirely due to
soft- and/or collinear {\bf gluon} exchange.
The infrared finite piece 
$d\Gamma_{77}^{(1,b),virt}/(ds_1 \, ds_2)$ can be written as
\begin{eqnarray}
&& \frac{d\Gamma_{77}^{(1,b),virt}}{ds_1 \, ds_2} = \frac{\alpha^2 \,  
\bar{m}_b^2(\mu) \, m_b^3 \, |C_{7,eff}(\mu)|^2 \, G_F^2 \,
  |V_{tb} V_{ts}^*|^2 \,  Q_d^2}{1024 \, \pi^5} \, \times  \nonumber \\
&&\hspace{2cm} \frac{\alpha_s}{4\pi} \, C_F \,  \, \left( 
\frac{-4 \, r_0 \, (1-s_1-s_2)}{(1-s_1)^2 \, s_1 \, (1-s_2)^2 \, s_2} \, \log \frac{\mu}{m_b} +
\frac{\sum_{i=1}^{20} v_i}{3 \, (1-s_1)^3 \, s_1 \, (1-s_2)^3 \, s_2} \right)
\label{virtualb} 
\end{eqnarray}
where the individual quantites $v_1,\ldots,v_{20}$ are relegated to
Appendix \ref{appendixvirt}.
\section{Bremsstrahlung corrections}
\label{sec:bremsstrahlung}
We now turn to the calculation of the bremsstrahlung QCD corrections, i.e. to
the contributions of order $\alpha_s$ with four particles in the
final state.
Before going into details, we mention that the kinematical range of
the variables $s_1$ and $s_2$ defined in
eq. (\ref{kinematicvariables}) is given  in this case
by $0 \le s_1 \le 1 \, ; \, 0 \le s_2 \le 1$. Nevertheless, we consider
in this paper only the range 
which is also accessible to the three-body decay $b \to s \gamma
\gamma$, i.e., $0 \le s_1 \le 1 \, ; \, 0 \le s_2 \le 1-s_1$ or, more
precisely, by its
restricted version specified in eq. (\ref{kinematical_cuts}).

\begin{figure}[h]
\begin{center}
\includegraphics[width=0.9\textwidth]{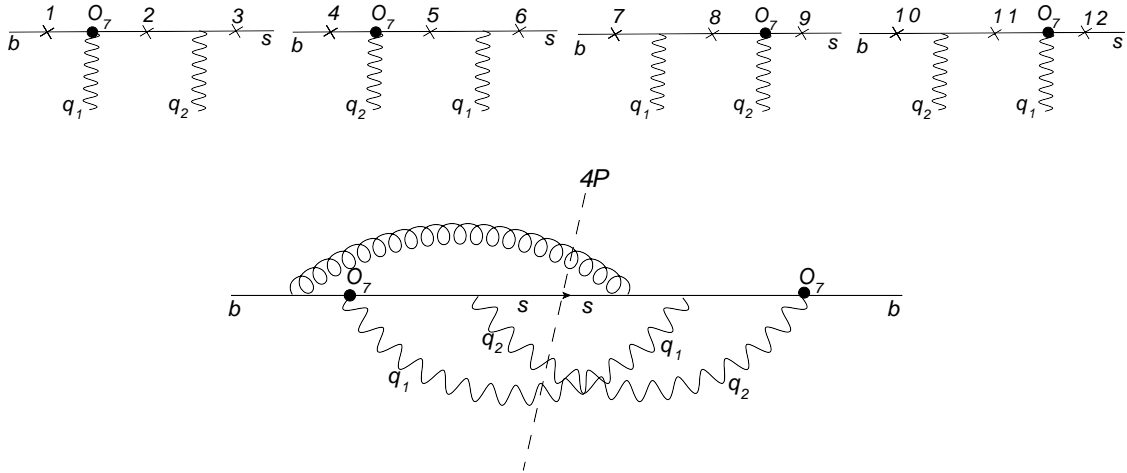}
\caption{On the first line the diagrams defining the
gluon-bremsstrahlung corrections to $b \to s \gamma \gamma$ are shown
at the amplitude level. The crosses in the graphs stand
for the possible emission places of the gluon.
On the second line the contribution to the decay width corresponding
to the interference of  diagram 1 with diagram 6 is illustrated.}
\label{fig:amplitudebrems}
\end{center}
\end{figure}

 The diagrams defining the bremsstrahlung corrections at the
amplitude level are shown in the first line of
Fig. \ref{fig:amplitudebrems}. 
The amplitude squared, needed to get the (double differential) decay width, can be written
as a sum of interferences of the different diagrams on the first line in
Fig. \ref{fig:amplitudebrems}. One such interference is shown on the
second line of the same figure. The four particle final state is
described by five independent kinematical variables.
In a first attempt we worked out the decay width by keeping $s_1$ and $s_2$
differential and integrating over the three remaining
variables. Proceeding in this  way, we found that the infrared- and
collinear singularities in the bremsstrahlung spectrum do not cancel
when adding the virtual corrections. The sum still contains $1/\epsilon$-poles,
but no $1/\epsilon^2$-poles.
While, as already mentioned
in section \ref{sec:virtual}, the only source of the
singularities in the virtual corrections in our restricted range of
$s_1$ and $s_2$ are due to soft {\bf gluon}-emission and/or
collinear emission of {\bf gluons} from the $s$-quark, we found after
analyzing the bremsstrahlung kinematics more carefully that there are situations
where one of the {\bf photons} can become collinear with the
$s$-quark. This is the reason why there is no cancellation of singularities
when combining virtual- and bremsstrahlung corrections. Realizing that
for (formally) zero hadronic mass of the $(s,g)$-system collinear
photon emission is kinematically impossible, led us to the idea that
we should first look at the triple differential decay width 
$d\Gamma_{77}/(ds_1 ds_2 ds_3)$, where $s_3=(p_s+p_g)^2/m_b^2$ is the normalized hadronic
mass squared. Our conjecture was that the double differential decay width,
based on the triple differential decay width in which only the leading
power terms w.r.t. $s_3$ are retained, should lead to a
finite result when combined with the virtual corrections.

We therefore worked out the leading power of this quantity w.r.t
$s_3$, denoting it by $d\Gamma_{77}^{\rm leading \, power}/(ds_1 ds_2
ds_3)$. The leading power, which is of  order $1/s_3$ (modified by
epsilontic dimensional regulators), is supposed to be a good
approximation for low values of the
hadronic mass. An approximation to the double-differential decay width
$d\Gamma^{(1),brems}_{77}/(ds_1 ds_2)$ due to gluon bremsstrahlung
corrections is then obtained by integrating \\ 
$d\Gamma_{77}^{\rm leading
  \, power}/(ds_1 ds_2 ds_3)$
over $s_3$, which runs in the range $s_3 \in [0,s_1 \cdot s_2]$. 
The
approximation is obviously accurate for small values of
$s_1 \cdot s_2$. As $s_1 \cdot s_2$ is at most $1/4$, the
approximation is  expected not to be bad in the full region of $s_1$
and $s_2$ considered in this paper. The technical details of the
calculation of the leading power w.r.t. $s_3$ in  the triple differential decay
width are illustrated in section \ref{subsec:bremsstrahlung}
 for the interference of diagram~1
with diagram~6, as shown in the second line of Fig. \ref{fig:amplitudebrems}.  

Indeed, we find that the infrared- and collinear singularities cancel when
combining the approximated version of $d\Gamma^{(1),brems}_{77}/(ds_1 ds_2)$
with the virtual corrections $d\Gamma^{(1),virt}_{77}/(ds_1 ds_2)$.

When going beyond this approximation other concepts, which go beyond
perturbation theory, like parton fragmentation
functions of a quark or a gluon into a photon, 
are needed \cite{Kapustin:1995fk}. We do not enter this issue in this paper.

The result of combined virtual- and bremsstrahlung corrections is
explicitly presented in the next section. 
\section{Final result for the decay width at order $\alpha_s$}\label{sec:combination}
The complete order $\alpha_s$ correction to the
double differential decay width
$d\Gamma_{77}/(ds_1 \, ds_2)$ is obtained by
adding the renormalized virtual corrections from section
\ref{sec:virtual} and the bremsstrahlung corrections discussed in
section \ref{sec:bremsstrahlung}. 
Explicitly we get
\begin{eqnarray}
&& \frac{d\Gamma_{77}^{(1)}}{ds_1 \, ds_2} = 
\frac{\alpha^2 \, \bar{m}_b^2(\mu) \, m_{b}^3 \, |C_{7,eff}(\mu)|^2 \, G_F^2 \,
  |V_{tb} V_{ts}^*|^2 \,  Q_d^2}{1024 \, \pi^5} \times \nonumber \\
&& \hspace{2cm} \frac{\alpha_s}{4\pi} \, C_F \,  \, 
\left[
\frac{-4 \, r_0 \, (1-s_1-s_2)}{(1-s_1)^2 \, s_1 \, (1-s_2)^2 \, s_2} \, \log \frac{\mu}{m_b} +f
\right]  \, ,
\label{total}
\end{eqnarray}
where $f$ is decomposed as
\begin{eqnarray}
&& f=\frac{\left(1-s_1-s_2\right) \left(f_1+f_2+f_3+f_4+f_5+f_6
   +f_7+f_8+f_9+f_{15}+
   f_{16}+f_{17}\right)
   }{3
   \left(1-s_1\right)^3 s_1
   \left(1-s_2\right)^3 s_2} \nonumber \\
&& \hspace{1cm} +\frac{f_{10}+f_{11}+
f_{12}+f_{13}+f_{14}}{3 \left(1-s_1\right)^3 s_1
   \left(1-s_2\right)^3 s_2}
\end{eqnarray}
The individual quantities $f_1,\ldots,f_{17}$ read
\begin{eqnarray}
\nonumber f_1&=&-16 \pi ^2 s_2^3 s_1^5-16\pi ^2 s_2^5 s_1^3+
48 \pi ^2 s_2^2 s_1^5+48 \pi ^2 s_2^5 s_1^2-48 \pi ^2 s_2 s_1^5-48 \pi ^2 s_2^5 s_1+
\\&&\nonumber
16 \pi ^2 s_1^5+16 \pi ^2 s_2^5-168 s_1^3-168 s_2^3+
\left(1104+160 \pi ^2\right) s_2^4 s_1^4+
   \\&&\nonumber
 \left(-3360-400 \pi ^2\right) s_2^3
   s_1^4+\left(3432+304 \pi ^2\right) s_2^2
   s_1^4+\left(-1296-64 \pi ^2\right) s_2
   s_1^4+
   \\&&\nonumber
   \left(120-16 \pi ^2\right) s_1^4+\left(-3360-400 \pi
   ^2\right) s_2^4 s_1^3+\left(10416+1152 \pi
   ^2\right) s_2^3 s_1^3+
    \\&&\nonumber
   \left(-10872-1056 \pi
   ^2\right) s_2^2 s_1^3+\left(3984+368 \pi
   ^2\right) s_2 s_1^3+\left(3432+304 \pi ^2\right)
   s_2^4 s_1^2+
    \\&&\nonumber
   \left(-10872-1056 \pi ^2\right)
   s_2^3 s_1^2+\left(12096+1088 \pi ^2\right)
   s_2^2 s_1^2+
    \\&&\nonumber
   \left(-4872-448 \pi ^2\right)
   s_2 s_1^2+\left(216+16 \pi ^2\right)
   s_1^2+\left(-1296-64
   \pi ^2\right) s_2^4 s_1+
    \\&&\nonumber
   \left(3984+368 \pi
   ^2\right) s_2^3 s_1+
   \left(-4872-448 \pi
   ^2\right) s_2^2 s_1+\left(2352+224 \pi
   ^2\right) s_2 s_1+
    \\&&\nonumber
   \left(-168-16 \pi
   ^2\right) s_1+\left(120-16
   \pi ^2\right) s_2^4+\left(216+16
   \pi ^2\right) s_2^2+\left(-168-16 \pi
   ^2\right) s_2
   \end{eqnarray}
   \begin{eqnarray}
   \nonumber 
f_2&=&   48 s_2\left(1-s_1\right) \left(1-s_2\right){}^2
    \left(6 s_2 s_1^3-6 s_1^3-11 s_2
   s_1^2+15 s_1^2+3 s_2 s_1-9 s_1+2\right)\times\\
\nonumber 
   && \log \left(1-s_1\right)
   \end{eqnarray}
   \begin{eqnarray}
\nonumber f_3 &=&   24 \left(1-s_1\right) \left(1-s_2\right)
   \left(30 s_2^3 s_1^3-64 s_2^2 s_1^3+41 s_2
   s_1^3-7 s_1^3-60 s_2^3 s_1^2+128 s_2^2
   s_1^2-\right.
   \\&&\left. \nonumber
   82 s_2 s_1^2+37 s_2^3 s_1-78 s_2^2
   s_1+76 s_2 s_1-7 s_1-7 s_2^3-7 s_2\right)
   \log \left(s_1\right)
   \end{eqnarray}
   \begin{eqnarray}
   \nonumber f_4 &=&-48 \left(1-s_1\right) \left(1-s_2\right)
   \left(s_2^2 s_1^4-s_2 s_1^4-5 s_2^3 s_1^3+9
   s_2^2 s_1^3-5 s_2 s_1^3+s_1^3+9 s_2^3
   s_1^2-\right.
   \\&& \left.  \nonumber
   20 s_2^2 s_1^2+13 s_2 s_1^2-5 s_2^3
   s_1+12 s_2^2 s_1-12 s_2
   s_1+s_1+s_2^3+s_2\right) \log
   ^2\left(s_1\right)
\end{eqnarray}
   \begin{eqnarray}
   \nonumber f_7 &=&
    96 \left(1-s_1\right) \left(1-s_2\right)
   \left(6 s_2^3 s_1^3-12 s_2^2 s_1^3+7 s_2
   s_1^3-s_1^3-12 s_2^3 s_1^2+24 s_2^2
   s_1^2-
   \right.
   \\&& \left. \nonumber
   14 s_2 s_1^2+7 s_2^3 s_1-14 s_2^2
   s_1+12 s_2 s_1-s_1-s_2^3-s_2\right) \log
   \left(s_1\right) \log \left(s_2\right)
   \end{eqnarray}
   \begin{eqnarray}
\nonumber f_9 &=&
   -96 \left(1-s_1\right) \left(1-s_2\right)
    \left(6 s_2^3
   s_1^3-12 s_2^2 s_1^3+7 s_2 s_1^3-s_1^3-12
   s_2^3 s_1^2+24 s_2^2 s_1^2-
\right.
   \\&& \left. \nonumber
      14 s_2 s_1^2+7
   s_2^3 s_1-14 s_2^2 s_1+12 s_2
   s_1-s_1-s_2^3-s_2\right) \log
   \left(s_1+s_2\right)
   \end{eqnarray}
   \begin{eqnarray}
 \nonumber f_{10} &=&
   96 \left(1-s_1\right) \left(1-s_2\right){}^2
   \left(s_2 s_1^5-s_1^5+2 s_2^2 s_1^4-5 s_2
   s_1^4+3 s_1^4+s_2^3 s_1^3-5 s_2^2 s_1^3+
\right.
   \\&& \left.  \nonumber 
   8 s_2 s_1^3-2 s_1^3-s_2^3 s_1^2+4 s_2^2
   s_1^2-4 s_2 s_1^2+s_1^2-4 s_2^2 s_1+3 s_2
   s_1-s_1-
\right.
   \\&& \left.  \nonumber    
   s_2^2+s_2\right) \log
   \left(1-s_1\right) \log
   \left(s_1+s_2\right)
      \end{eqnarray}
   \begin{eqnarray}
  \nonumber f_{11} &=&
   -96 \left(1-s_1\right) \left(1-s_2\right)
   \left(s_2^2 s_1^5-s_2 s_1^5-10 s_2^3
   s_1^4+19 s_2^2 s_1^4-11 s_2 s_1^4+2
   s_1^4-
\right.
   \\&& \left.  \nonumber       
   11 s_2^4 s_1^3+53 s_2^3 s_1^3-77
   s_2^2 s_1^3+41 s_2 s_1^3-2 s_1^3+21 s_2^4
   s_1^2-76 s_2^3 s_1^2+94 s_2^2 s_1^2-
\right.
   \\&& \left. \nonumber           
   49 s_2
   s_1^2+2 s_1^2-11 s_2^4 s_1+38 s_2^3 s_1-46
   s_2^2 s_1+25 s_2 s_1-2
   s_1+s_2^4-s_2^3+
\right.
   \\&& \left.   \nonumber            
   s_2^2-s_2\right) \log
   \left(s_1\right) \log \left(s_1+s_2\right)
      \end{eqnarray}
   \begin{eqnarray}
    \nonumber f_{14} &=&
   48 \left(1-s_1\right) \left(1-s_2\right)
   \left(s_2^2 s_1^5-s_2 s_1^5-21 s_2^3
   s_1^4+40 s_2^2 s_1^4-22 s_2 s_1^4+3
   s_1^4-
\right.
   \\&& \left. \nonumber                 
   21 s_2^4 s_1^3+106 s_2^3 s_1^3-153
   s_2^2 s_1^3+79 s_2 s_1^3-3 s_1^3+s_2^5
   s_1^2+40 s_2^4 s_1^2-153 s_2^3 s_1^2+
\right.
   \\&& \left. \nonumber                    
   188 s_2^2 s_1^2-95 s_2 s_1^2+3 s_1^2-s_2^5
   s_1-22 s_2^4 s_1+79 s_2^3 s_1-95 s_2^2
   s_1+50 s_2 s_1-
\right.
   \\&& \left. \nonumber                        
   3 s_1+3 s_2^4-3 s_2^3+3
   s_2^2-3 s_2\right) \log
   ^2\left(s_1+s_2\right)
   \end{eqnarray}
   \begin{eqnarray}
\nonumber f_{15} &=&
   96 s_1 \left(1-s_2\right){}^2 \left(s_2
   s_1^4-s_1^4+s_2^2 s_1^3-4 s_2 s_1^3+3
   s_1^3-5 s_2^2 s_1^2+8 s_2 s_1^2-2 s_1^2+
\right.
   \\&& \left. \nonumber                           
   7 s_2^2 s_1-11 s_2 s_1+s_1-2 s_2^2+5
   s_2-1\right) \text{Li}_2\left(s_1\right)
      \end{eqnarray}
   \begin{eqnarray}
 \nonumber f_{16} &=&96 \left(1-s_1\right) \left(1-s_2\right)
   \left(s_2^2 s_1^4-2 s_2 s_1^4+s_1^4+8 s_2^3
   s_1^3-17 s_2^2 s_1^3+12 s_2 s_1^3-
\right.
   \\&& \left. \nonumber                              
   3 s_1^3+s_2^4 s_1^2-17 s_2^3 s_1^2+32 s_2^2
   s_1^2-20 s_2 s_1^2-2 s_2^4 s_1+12 s_2^3
   s_1-20 s_2^2 s_1+
\right.
   \\&& \left. \nonumber                                 
   20 s_2 s_1-2 s_1+s_2^4-3
   s_2^3-2 s_2\right)
   \text{Li}_2\left(1-s_1-s_2\right)
\end{eqnarray}
\begin{eqnarray}
 && \nonumber f_5=f_2(s_1\leftrightarrow s_2)\quad \quad
 f_6=f_3(s_1\leftrightarrow s_2)\quad \quad
 f_8=f_4(s_1\leftrightarrow s_2)\\
\nonumber &&f_{12}=f_{10}(s_1\leftrightarrow s_2)\quad \quad
 f_{13}=f_{11}(s_1\leftrightarrow s_2)\quad \quad
 f_{17}=f_{15}(s_1\leftrightarrow s_2)
\end{eqnarray}
The order $\alpha_s$ correction $d\Gamma_{77}^{(1)}/(ds_1 ds_2)$ in
Eq. (\ref{total}) to the double differential decay width
for $b \to X_s \gamma \gamma$ is the
main result of our paper.

\section{Some numerical illustrations}\label{sec:numerics}
In the previous sections we calculated the virtual- and bremsstrahlung
QCD corrections which were the missing ingredient in order to obtain the
$({\cal O}_7,{\cal O}_7)$ contribution to the double differential decay
width for $\bar{B} \to X_s \gamma \gamma$ at NLL precision.
The Wilson coefficient $C_{7,eff}(\mu)$ at the low scale $(\mu \sim m_b)$
which is needed up to order $\alpha_s$, i.e.,
\be
C_{7,eff}(\mu) = C_{7,eff}^{0}(\mu) + \frac{\alpha_s(\mu)}{4\pi} \, C_{7,eff}^{1}(\mu)
\label{wilsonexpand}
\ee
is known for a long time (see ref. \cite{Chetyrkin:1996vx} and
references therein).
Numerical values for
the input parameters and for this Wilson coefficient at various values
for the scale $\mu$, together with the
numerical values of $\alpha_s(\mu)$, are given in Table
\ref{tab:input} and Table \ref{tab:wilson}, respectively.
\begin{table}[h]
\centering \vspace{0.8cm}
\begin{tabular}{|c|c|}

 \hline
 Parameter & Value \\   \hline \hline

$m_{b}$& $4.8$~GeV   \\ \hline

$m_{t}$&$175$~GeV    \\ \hline

$m_{W}$&$80.4$~GeV    \\ \hline

$m_{Z}$&$91.19$~GeV   \\ \hline

$G_{F}$&$1.16637\times10^{-5}$~\text{GeV}$^{-2}$  \\ \hline

$V_{tb} V_{ts}^* $&$0.04$    \\  \hline

${\alpha}^{-1}$&$137$    \\ \hline

${\alpha_{s}(M_{Z})}$&$0.119$      \\ \hline \hline

\end{tabular}
\caption{Values of the relevant input parameters} 
\label{tab:input}
\end{table}
\begin{table}[h]
\centering \vspace{0.8cm}
\begin{tabular}{|c|c|c|c|}
\hline
  &  $\alpha_s(\mu)$ & $C_{7,eff}^{0}(\mu)$ & $C_{7,eff}^{1}(\mu)$
\\   \hline \hline

$\mu= m_W$  & $0.1213$ & $-0.1957$ & $-2.3835$ \\ \hline 

$\mu=2 \, m_{b}$  & $0.1818$ & $-0.2796$ & $-0.1788$ \\ \hline 

$\mu=m_{b}$  & $0.2175$ & $-0.3142$ &  $0.4728$   \\ \hline

$\mu=m_{b}/2$  & $0.2714$ & $-0.3556$ & $1.0794$    \\ \hline \hline

\end{tabular}
  \caption{$\alpha_s(\mu)$ and the Wilson coefficient $C_{7,eff}(\mu)$ at
    different values of the scale $\mu$} 
\label{tab:wilson}
\end{table}
The NLL prediction reads
\be
\frac{d\Gamma_{77}}{ds_1 ds_2} = 
\frac{d\Gamma_{77}^{(0)}}{ds_1 ds_2} + \frac{d\Gamma_{77}^{(1)}}{ds_1 ds_2}
\label{total77}
\ee  
where the first- and second term of the r.h.s. are given in eqs. (\ref{treezero})
and (\ref{total}), respectively. 

\begin{figure}[h]
\begin{center}
\includegraphics[width=\textwidth]{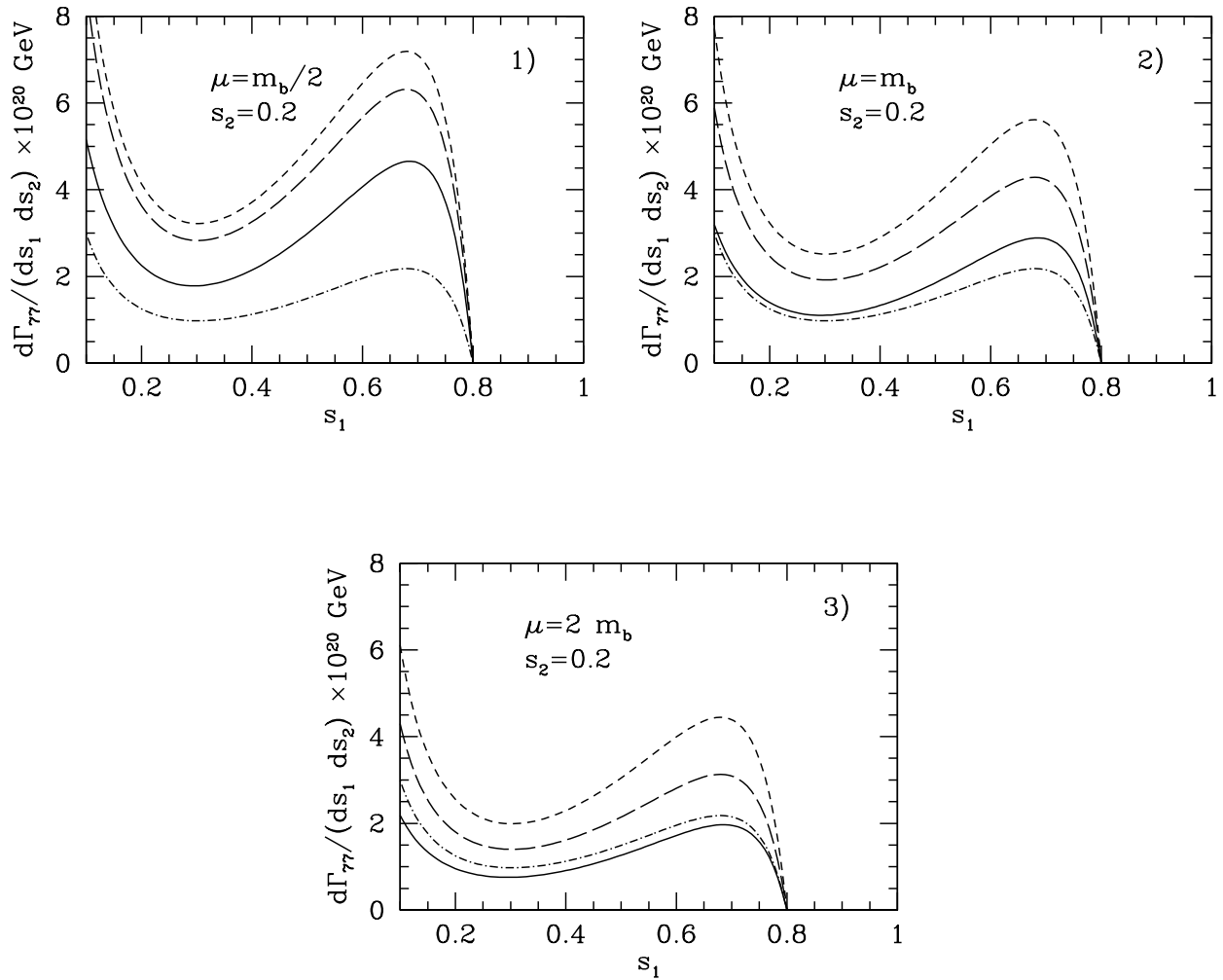}
\caption{Double differential decay width $d\Gamma_{77}/(ds_1 ds_2)$ 
as a function of $s_1$ for $s_2$ fixed at $s_2=0.2$. 
The dash-dotted, the short-dashed and the solid line shows the result 
when neglecting QCD-effects, the LL result and the NLL result,
respectively. The long-dashed line represents the (partial) NLL result
in which the virtual- and bremsstrahlung corrections worked out in this
paper
are switched off (see text for more details).
In the frames 1), 2) and 3) the renormalization scale is chosen to
  be $\mu=m_b/2$, $\mu=m_b$ and $\mu=2 \, m_b$, respectively.}
\label{fig:results}
\end{center}
\end{figure}

To illustrate our results, we first rewrite the $\overline{\mbox{MS}}$ mass
$\bar{m}_b(\mu)$ in eq. (\ref{total77}) in terms of the pole mass
$m_b$,
 using the one-loop relation
\[
\bar{m}_b(\mu) = m_b \, \left[ 1 - \frac{\alpha_s(\mu)}{4 \pi} \, \left( 8 \log
  \frac{\mu}{m_b} + \frac{16}{3} \right)
\right] \, .
\]
We then insert $C_{7,eff}(\mu)$ in the expanded form
(\ref{wilsonexpand}) and expand the resulting expression for
$d\Gamma_{77}/(ds_1 ds_2)$ w.r.t. $\alpha_s$, discarding terms of
order $\alpha_s^2$. This defines the NLL result. The
corresponding LL result is obtained by also discarding the order
$\alpha_s^1$ term. In Fig. \ref{fig:results} the LL- and the NLL-
result is shown by the short-dashed- and the solid line,
respectively. 

In our procedure the NLL corrections have three sources: 
(a) $\alpha_s$ corrections to the Wilson coefficient $C_{7,eff}(\mu)$,
(b) expressing $\bar{m}_b(\mu)$ in terms of the pole mass $m_b$ and
(c) virtual- and real- order $\alpha_s$ corrections to the matrix
elements. To illustrate the effect of source (c), which is worked out
for the first time in this paper, we show in Fig. \ref{fig:results}
(by the long-dashed line) the (partial) NLL result in which 
source (c) is switched off. 
We conclude that the effect (c) is roughly of equal
importance as the combined effects of (a) and (b).

For completeness we
show in this figure (by the dash-dotted line) also the result when QCD
is completely switched off, which amounts to put $\mu=m_W$ in the LL
result.

From Fig. \ref{fig:results} we see that the NLL results are
substantially smaller (typically by $50\%$ or slightly more) than
those at LL precision, which is also the case when choosing other
values for $s_2$. 
\newpage

%
%
%

\section{Technical details about our calculations}
\label{sec:technical}
We first describe the general setup of our calculations and then
discuss in the subsections \ref{subsec:virtual} and \ref{subsec:bremsstrahlung}
the calculation of the virtual- and the bremsstrahlung corrections for
the interference diagrams shown in the last lines of
Fig. \ref{fig:amplitudevirtual} and
Fig. \ref{fig:amplitudebrems}, respectively.

The starting point is the general expression for the total decay width of the
massive $b$ quark with momentum $p_b$ decaying into $3\le n\le4$
massless final-state particles with momenta $k_i$,
\begin{align}\label{general_phasespace}
  \Gamma_{1\to n} &= \frac{1}{2m_b}
\left(\prod_{i=1}^n\int\!\frac{d^{d-1}k_i}{(2\pi)^{d-1}2E_i}\right)(2\pi)^d
    \,\delta^{(d)}\left(p_b-\sum_{i=1}^nk_i\right)\,|M_n|^2\nn\\[1mm]
  &= \frac{1}{2m_b}(2\pi)^n
    \left(\prod_{i=1}^{n-1}\int\!\frac{d^dk_i}{(2\pi)^d}\,\delta(k_i^2)\,
      \theta(k_i^0)\right) \nn\\[1mm]
  &\qquad\times\delta\left(\left(p_b-\sum_{i=1}^{n-1}k_i\right)^2\right)\,
    \theta\left(p_b^0-\sum_{i=1}^{n-1}k_i^0\right)\,|M_n|^2\, ,
\end{align}
where the squared Feynman amplitude $|M_n|^2$ is always understood to be
summed
over final spin-, polarization- and color states, and averaged over the
spin
directions and colors of the decaying $b$ quark. It also
includes a
factor of $1/2$ for the two identical particles in the final state,
i.e. the photons.
Furthermore, $d=4-2\epsilon$ denotes the space-time dimension that we use
to
regulate the ultraviolet, infrared and collinear singularities.

The double differential decay rate $d\Gamma_{77}/(ds_1 ds_2)$ is obtained from
eq. (\ref{general_phasespace}) by multiplying the integrand on the r.h.s.
with the delta functions 
$\delta \left(s_1-(p_b-q_1)^2/m_b^2 \right )$ and
$\delta \left(s_2-(p_b-q_2)^2/m_b^2 \right )$ \cite{Anastasiou:2002qz},
where $p_b$ and $q_1$, $q_2$ denote the four momenta of the $b$ quark and 
the photons, respectively. For the bremsstrahlung corrections, as
mentioned in section \ref{sec:bremsstrahlung}, we need to consider
also the triple differential decay width $d\Gamma_{77}/(ds_1 ds_2 ds_3)$, 
where $s_3=(p_b-q_1-q_2)^2/m_b^2$ is the normalized hadronic mass squared. 
The triple differential decay width is obtained by multiplying the integrand with
the additional delta-function
$\delta \left(s_3-(p_b-q_1-q_2)^2/m_b^2\right)$.
Finally, the delta functions just mentioned and all of the delta
functions present in eq. (\ref{general_phasespace}) can be rewritten as differences 
of propagators as follows \cite{Anastasiou:2002yz,Anastasiou:2003ds},
\be \label{deltoprops}
\delta\left( q^2  - m^2 \right) = \frac{1}{2 \pi i} \left( 
\frac{1}{ q^2  - m^2 -i 0} - \frac{1}{q^2  - m^2 +i 0} \right) \, .
\ee
In this step the phase-space integrations are converted into loop integrations
(which can be combined with possible loop integrations already present
in $|M_n|^2$).
By subsequently doing tensor reductions, the (differential) decay width can be
written as a linear combination of scalar integrals. 
The systematic Laporta algorithm \cite{Laporta:2001dd}, based
on integration-by-part techniques first proposed in
\cite{Tkachov:1981wb,Chetyrkin:1981qh}, can then be applied to reduce
the scalar integrals to a small number of simpler integrals, usually
referred to as the master integrals (MIs). 
For this reduction we used the AIR and FIRE implementations 
\cite{Anastasiou:2004vj,Smirnov:2008iw} of the Laporta algorithm.
After the reduction process, it usually happens that some MIs contain
propagators which were introduced via (\ref{deltoprops}) with zero or
negative power. In this case the $\pm i0$ prescription becomes
irrelevant and as a consequence
these MIs are zero. In the remaining MIs we convert
the propagators introduced via (\ref{deltoprops}) back to
delta-functions. Thus, we are left with phase-space MIs (which can contain
loop integrations as well).
The final task is then to calculate these MIs, i.e. to perform possible loop
integrations together with the phase-space integrations.

Very often we had to deal with MIs which we were not able to evaluate by
direct integration of their integral representation in terms of
Feynman parameters and/or phase-space parameters.
A powerful tool to be used in these cases is the differential
equation method
\cite{Anastasiou:2003ds,Anastasiou:2002yz,Remiddi:1997ny,Argeri}. 
The goal of this method is to employ the output of the reduction procedure for a given
topology to build differential equations which are satisfied by the MIs
of that topology. In our case, we consider differential equations w.r.t.
$s_1$ and $s_2$ and also w.r.t. $s_3$ for case of bremsstrahlung corrections.  
With these methods we were able to obtain analytic
expressions for all master integrals appearing in the calculation of
the various diagrams. 

\subsection{Details about the calculation of virtual corrections}
\label{subsec:virtual}
To illustrate our methods for the virtual corrections, we take as an
example the interference diagram shown on the last line of Figure
\ref{fig:amplitudevirtual}. In this
case we have five MIs. Four of them can be solved  by means of direct integration on 
Feynman parameters. To calculate the last one (called $P_{1111}$),
we solve the differential equations w.r.t. $s_1$ and $s_2$ and get
the solution which we denote as $Q_{1111}$. At this level $Q_{1111}$ contains
integration constants (which are not fixed by the differential
equations). 
To get the integration constants, we proceed in the following way.
Using Feynman parametrization for the loop integral, we write the MI 
$P_{1111}$ as
\be
P_{1111}=s_1^{-\epsilon} s_2^{-\epsilon} (1-s_1-s_2)^{-\epsilon}
\int_0^1 g_0(s_1,s_2,\epsilon,u,v,y)\, du \, dv \,dy \, ,
\ee
where $u$, $v$ and $y$ are Feynman parameters (all of them running  from 0 to 1).
The factor $s_1^{-\epsilon} s_2^{-\epsilon} (1-s_1-s_2)^{-\epsilon}$ is coming 
from the phase-space (see eq. (\ref{phasespace3}) in 
Appendix \ref{subsec:phasespace3}). 
One then can put $s_2=0$ in $g_0(s_1,s_2,\epsilon,u,v,y)$
and integrate on the Feynman parameters, which
defines the new function 
\be
g_1(s_1,\epsilon)=\int_0^1 g_0(s_1,0,\epsilon,u,v,y) \, du \, dv \, dy
\, .
\ee
We managed to work out the leading term of the expansion of 
$g_1(s_1,\epsilon)$ on $s_1$ around zero, which is proportional to $s_1^{-2}$. 
From this, we immediately get the leading term of the expansion of  $P_{1111}$
on $s_2$ and $s_1$ (which is proportional to $s_2^0/s_1^2$).
Comparing the result of this calculation with the corresponding expansion 
of the solution $Q_{1111}$ of the differential equations,
we could determine all integration constants. 

\subsection{Details about the calculation of bremsstrahlung
  corrections}
\label{subsec:bremsstrahlung}
To illustrate our methods for the bremsstrahlung corrections, we take
as an example the interference diagram shown on the last line of Figure 
\ref{fig:amplitudebrems}. In this case we obtain three MIs, denoted by
$P_{00}$, $P_{10}$ and $P_{11}$. Writing the diagram as a linear
combination of the MIs, we see that the leading power (w.r.t. $s_3$) of 
all three coefficients (in front of the MIs) is of the order $1/s_3$.  
Keeping in mind that we are taking into account only terms proportional 
to the leading power in $s_3$ at the level of the triple differential
decay width
(as extensively discussed in section
\ref{sec:bremsstrahlung}), it is sufficient to work out
the MIs to zero-th power in $s_3$, including the epsilontic regulator, i.e.,
$s_3^{0-n \epsilon}$ (in our case only $n=1$ and $n=2$ occur). 

The simplest MI, $P_{00}$, which corresponds to the pure phase-space
(see eq. (\ref{phasespace4}) in Appendix \ref{subsec:phasespace4}), 
can be easily 
solved  by means of direct integration. 
For  $P_{10}$  the solution of the differential equation
w.r.t. $s_3$ can be represented in the limit $s_3\to 0$ in the form
\be
P_{10}=a_1(s_1,s_2,\epsilon) \, s_3^{-\epsilon}+a_2(s_1,s_2,\epsilon) \,
s_3^{-2\epsilon} \, ,
\ee
where the function $a_1(s_1,s_2,\epsilon)$ is fully determined. 
To find the function $a_2(s_1,s_2,\epsilon)$,
we use differential equations w.r.t. $s_1$ and $s_2$.
In this way, we could find $a_2(s_1,s_2,\epsilon)$ up to integration constants.
To determine these constants, we managed to calculate  the
MI for specific values of $s_1$, $s_2$ and $s_3 \to 0$.
In the same way we also calculated the MI $P_{11}$.
\section{Summary}\label{sec:summary}
In the present work we calculated the set of the $O(\alpha_s)$
corrections to the decay process $\bar{B} \to X_s \gamma \gamma$
originating from diagrams involving the electromagnetic
dipole operator ${\cal O}_7$.
To perform this calculation it is necessary to work out diagrams
with three particles ($s$-quark and two photons) and four
particles ($s$-quark, two photons and a gluon) in the final state.
From the technical point of view, the calculation was made possible
by the use of the Laporta Algorithm  to identify the
needed Master Integrals and by applying the differential equation method to
solve the Master Integrals.
When calculating the bremsstrahlung corrections, we take into account
only terms proportional to the leading power of the hadronic mass.
We find that the infrared and collinear singularities cancel when
combining the above mentioned approximated version of bremsstrahlung
corrections with the virtual corrections.
The numerical impact of the NLL corrections is not small: for
$d\Gamma_{77}/(ds_1 \, ds_2)$ the NLL results are approximately
50\% smaller than those at LL precision.

\section*{\normalsize Acknowledgments}
\vspace*{-2mm}
C.G. and A.K. were partially supported by the Swiss National Foundation and by
the Helmholz Association through 
funds provided to the virtual institute ``Spin and strong QCD''
(VH-VI-231). The Albert Einstein Center for Fundamental Physics
(Bern), to which C.G. and A.K. are affiliated, is supported by the
``Innovations- und Kooperationsprojekt C-13 of the Schweizerische Universit\"atskonferenz SUK/CRUS''.

\noindent
H.A. and  A.Y. were supported by the Armenian State Committee of Science under
contract 11-1c014.

\noindent
Finally, we thank A.V. Smirnov for helpful email exchanges concerning his
program FIRE \cite{Smirnov:2008iw}.

\newpage

\appendix

\section{Explicit results for the functions $v_i$ defining the
virtual corrections}
\label{appendixvirt}
The functions $v_i$ appearing in eq. (\ref{virtualb}) read
\begin{eqnarray}
\nonumber v_1&=& \left(1-s_1-s_2\right) \left(
-16 \pi ^2 s_2^3 s_1^5-16 \pi ^2 s_2^5 s_1^3+48 \pi ^2 s_2^2 s_1^5+
   48\pi ^2 s_2^5 s_1^2-48 \pi ^2 s_2 s_1^5-
   \right.  \\&& \left.  \nonumber
   48\pi ^2 s_2^5 s_1+
   16 \pi ^2 s_1^5+16 \pi ^2 s_2^5+
   \left(2112-104 \pi
   ^2\right) s_2^4 s_1^4+\left(392 \pi
   ^2-6384\right) s_2^3 s_1^4+
   \right.  \\&& \left.  \nonumber
   \left(6672-532
   \pi ^2\right) s_2^2 s_1^4+\left(288 \pi
   ^2-2736\right) s_2 s_1^4+\left(336-60 \pi
   ^2\right) s_1^4+
   \right.  \\&& \left.  \nonumber
   \left(392 \pi ^2-6384\right) s_2^4
   s_1^3+\left(19584-1224 \pi ^2\right) s_2^3
   s_1^3+\left(1452 \pi ^2-20784\right) s_2^2
   s_1^3+
   \right.  \\&& \left.  \nonumber
   \left(7872-600 \pi ^2\right) s_2
   s_1^3+\left(44 \pi ^2-288\right) s_1^3+\left(6672-532 \pi
   ^2\right) s_2^4 s_1^2+
   \right.  \\&& \left.  \nonumber
   \left(1452 \pi
   ^2-20784\right) s_2^3
   s_1^2+\left(23904-1728 \pi ^2\right) s_2^2
   s_1^2+\left(740 \pi ^2-10128\right) s_2
   s_1^2+
   \right.  \\&& \left.  \nonumber
   \left(336-28 \pi ^2\right) s_1^2+\left(288 \pi
   ^2-2736\right) s_2^4 s_1+\left(7872-600 \pi
   ^2\right) s_2^3 s_1+
   \right.  \\&& \left.  \nonumber
   \left(740 \pi
   ^2-10128\right) s_2^2 s_1+\left(5376-392
   \pi ^2\right) s_2 s_1+\left(28 \pi
   ^2-384\right) s_1+
   \right.  \\&& \left.  \nonumber
   \left(336-60 \pi ^2\right)
   s_2^4+\left(44 \pi ^2-288\right)
   s_2^3+\left(336-28 \pi ^2\right)
   s_2^2+\left(28 \pi ^2-384\right) s_2\right)
   \end{eqnarray}
 \begin{eqnarray}
\nonumber v_2&=&   -96 \left(1-s_1\right) \left(1-s_2\right)
   \left(1-s_1-s_2\right) \left(3 s_2^3
   s_1^3-4 s_2^2 s_1^3+s_2 s_1^3-5 s_2^3
   s_1^2+7 s_2^2 s_1^2-
   \right.  \\&& \left.  \nonumber
   2 s_2 s_1^2-s_1^2+2
   s_2^3 s_1-3 s_2^2 s_1+3 s_2
   s_1-s_2^2\right) \log \left(s_1\right)
 \end{eqnarray}
 \begin{eqnarray}
\nonumber 
v_3&=& -24 \left(1-s_1\right) \left(1-s_2\right)
   \left(1-s_1-s_2\right) \left(2 s_2^2
   s_1^4-2 s_2 s_1^4-4 s_2^3 s_1^3+6 s_2^2
   s_1^3-
   \right.  \\&& \left.  \nonumber
   3 s_2 s_1^3+s_1^3+
   6 s_2^3 s_1^2-16
   s_2^2 s_1^2+12 s_2 s_1^2-3 s_2^3 s_1+10
   s_2^2 s_1-12 s_2 s_1+
   \right.  \\&& \left.  \nonumber
   s_1+s_2^3+s_2\right)
   \log ^2\left(s_1\right)
 \end{eqnarray}
 \begin{eqnarray}
\nonumber v_4&=&
48 \left(1-s_1\right) \left(1-s_2\right)
   \left(1-s_1-s_2\right) \left(6 s_2^3
   s_1^3-12 s_2^2 s_1^3+7 s_2 s_1^3-s_1^3-12
   s_2^3 s_1^2+
   \right.  \\&& \left.  \nonumber
   24 s_2^2 s_1^2-14 s_2 s_1^2+7
   s_2^3 s_1-14 s_2^2 s_1+12 s_2
   s_1-s_1-s_2^3-s_2\right) \log
   \left(s_1\right) \log \left(s_2\right)
   \end{eqnarray}
\begin{eqnarray}
\nonumber v_5&=&48 \left(1-s_1\right) \left(1-s_2\right)
   \left(1-s_1-s_2\right) \left(6 s_2^3
   s_1^3-12 s_2^2 s_1^3+7 s_2 s_1^3-s_1^3-12
   s_2^3 s_1^2+
   \right.  \\&& \left.  \nonumber
   24 s_2^2 s_1^2-14 s_2 s_1^2+7
   s_2^3 s_1-14 s_2^2 s_1+12 s_2
   s_1-s_1-s_2^3-s_2\right) \log
   \left(s_1\right) \times
    \\&&   \nonumber 
   \log\left(1-s_1-s_2\right)
\end{eqnarray}
\begin{eqnarray}
\nonumber v_6&=&
-96 \left(1-s_1\right){}^2
   \left(1-s_2\right){}^2 s_2 \left(s_1^4+2
   s_2 s_1^3-2 s_1^3+s_2^2 s_1^2-4 s_2
   s_1^2+s_1^2-2 s_2^2 s_1+
\right.  \\&& \left.  \nonumber   
   3 s_2 s_1-2
   s_1+s_2^2+1\right) \log \left(s_1\right)
   \log \left(s_1+s_2\right)\end{eqnarray}
\begin{eqnarray}
\nonumber v_7&=&
48 \left(1-s_1\right) \left(s_2-1\right){}^2
   s_2 \left(1-s_1-s_2\right) \left(6 s_2
   s_1^3-6 s_1^3-11 s_2 s_1^2+15 s_1^2+
   \right.  \\&& \left.  \nonumber
   3 s_2 s_1-9 s_1+2\right) \log \left(1-s_1\right)
   \end{eqnarray}
\begin{eqnarray}
\nonumber v_8&=&
96 \left(1-s_1\right) \left(s_2-1\right){}^2
   \left(s_2 s_1^5-s_1^5+2 s_2^2 s_1^4-5 s_2
   s_1^4+3 s_1^4+s_2^3 s_1^3-5 s_2^2 s_1^3+
   \right.  \\&& \left.  \nonumber
   8 s_2 s_1^3-2 s_1^3-s_2^3 s_1^2+4 s_2^2
   s_1^2-4 s_2 s_1^2+s_1^2-4 s_2^2 s_1+3 s_2
   s_1-s_1-s_2^2+s_2\right) \times
   \\&&   \nonumber
   \log\left(1-s_1\right) \log
   \left(s_1+s_2\right)
   \end{eqnarray}
\begin{eqnarray}
\nonumber v_{9}&=&48 \left(1-s_1\right) \left(1-s_2\right)
   \left(s_2^2 s_1^5-s_2 s_1^5-9 s_2^3
   s_1^4+16 s_2^2 s_1^4-8 s_2 s_1^4+s_1^4-9
   s_2^4 s_1^3+
 \right.  \\&& \left.  \nonumber   
   46 s_2^3 s_1^3-67 s_2^2
   s_1^3+35 s_2 s_1^3-s_1^3+s_2^5 s_1^2+16
   s_2^4 s_1^2-67 s_2^3 s_1^2+84 s_2^2
   s_1^2-
   \right.  \\&& \left.  \nonumber 
   43 s_2 s_1^2+s_1^2-s_2^5 s_1-8 s_2^4
   s_1+35 s_2^3 s_1-43 s_2^2 s_1+22 s_2
   s_1-s_1+s_2^4-s_2^3+
   \right.  \\&& \left.  \nonumber 
   s_2^2-s_2\right) \log
   ^2\left(s_1+s_2\right)
\end{eqnarray}
\begin{eqnarray}
\nonumber v_{10}&=&-96 \left(1-s_1\right) \left(1-s_2\right)
   \left(1-s_1-s_2\right) \left(s_2^2
   s_1^3-s_2 s_1^3+s_2^3 s_1^2-3 s_2^2 s_1^2+2
   s_2 s_1^2-
  \right.  \\&& \left.  \nonumber 
   s_1^2-s_2^3 s_1+2 s_2^2 s_1+s_2
   s_1-s_2^2\right) \log
   \left(1-s_1-s_2\right)
\end{eqnarray}
\begin{eqnarray}
\nonumber v_{11}&=&24 \left(1-s_1\right) \left(1-s_2\right)
   \left(1-s_1-s_2\right) \left(6 s_2^3
   s_1^3-12 s_2^2 s_1^3+7 s_2 s_1^3-s_1^3-12
   s_2^3 s_1^2+
   \right.  \\&& \left.  \nonumber
   24 s_2^2 s_1^2-14 s_2 s_1^2+7
   s_2^3 s_1-14 s_2^2 s_1+12 s_2
   s_1-s_1-s_2^3-s_2\right) \times\\&&\nonumber
   \log^2\left(1-s_1-s_2\right)
\end{eqnarray}
\begin{eqnarray}
\nonumber v_{12}&=&96 s_1 \left(1-s_2\right){}^2
   \left(1-s_1-s_2\right) \left(s_2
   s_1^4-s_1^4+s_2^2 s_1^3-4 s_2 s_1^3+3
   s_1^3-5 s_2^2 s_1^2+
   \right.  \\&& \left.  \nonumber
   8 s_2 s_1^2-2 s_1^2+7
   s_2^2 s_1-11 s_2 s_1+s_1-2 s_2^2+5
   s_2-1\right) \text{Li}_2\left(s_1\right)
\end{eqnarray}
\begin{eqnarray}
\nonumber v_{13}&=&96 \left(1-s_1\right) \left(1-s_2\right)
   \left(1-s_1-s_2\right) \left(s_2^2 s_1^4-2
   s_2 s_1^4+s_1^4+8 s_2^3 s_1^3-17 s_2^2
   s_1^3+
   \right.  \\&& \left.  \nonumber
   12 s_2 s_1^3-3 s_1^3+s_2^4 s_1^2-17
   s_2^3 s_1^2+32 s_2^2 s_1^2-20 s_2 s_1^2-2
   s_2^4 s_1+12 s_2^3 s_1-
   \right.  \\&& \left.  \nonumber
   20 s_2^2 s_1+20 s_2
   s_1-2 s_1+s_2^4-3 s_2^3-2 s_2\right)
   \text{Li}_2\left(1-s_1-s_2\right)
\end{eqnarray}
\begin{eqnarray}
 && \nonumber v_{14}=v_2(s_1\leftrightarrow s_2)\quad \quad
 v_{15}=v_3(s_1\leftrightarrow s_2)\quad \quad
 v_{16}=v_5(s_1\leftrightarrow s_2)\\
\nonumber &&v_{17}=v_{6}(s_1\leftrightarrow s_2)\quad \quad
 v_{18}=v_{7}(s_1\leftrightarrow s_2)\quad \quad
 v_{19}=v_{8}(s_1\leftrightarrow s_2)
 \quad \quad
 v_{20}=v_{12}(s_1\leftrightarrow s_2)\end{eqnarray}
\section{Relevant phase-space formulas}
\subsection{Double differential phase-space for the 3-particle final
  state}
\label{subsec:phasespace3}
The kinematical variables $s_1$ and $s_2$ are defined as
\begin{equation}
s_1=\frac{(p_b-q_1)^2}{m_b^2} \quad ; \quad s_2=\frac{(p_b-q_2)^2}{m_b^2}\, ,
\end{equation}
where $p_b$ and $q_i$  denote the four-momenta of the $b$-quark and
the photons, respectively. The kinematics of the process
$b \to s \gamma \gamma$ is fully described by 
$s_1$ and $s_2$. The formula for double differential
decay width is therefore free of additional phase-space integration
variables. It reads
\begin{equation}
\frac{d\Gamma_{1 \to 3}}{ds_1 \, ds_2} = \frac{1}{4} \,
\frac{(4\pi)^{-3+2\epsilon}}{\Gamma[2-2\epsilon]} \, m_b^{1-4\epsilon}
\, s_1^{-\epsilon} \, s_2^{-\epsilon} \, (1-s_1-s_2)^{-\epsilon} \, |M_3|^2 \, .
\label{phasespace3}
\end{equation}
The variables $s_1$ and $s_2$ vary in the range
\[
0 \le s_1 \le 1 \, ; \quad 0 \le s_2 \le 1-s_1 \, .
\]
\subsection{Triple differential phase-space for the 4-particle final state}
\label{subsec:phasespace4}
The kinematical variables $s_1$, $s_2$ and $s_3$ are defined as
\begin{equation}
s_1=\frac{(p_b-q_1)^2}{m_b^2} \quad ; \quad
s_2=\frac{(p_b-q_2)^2}{m_b^2} \quad ; \quad s_3=\frac{(p_s+k)^2}{m_b^2}
\end{equation}
where $p_b$, $p_s$, $k$ and $q_i$  denote the four-momenta of the
$b$-quark, the $s$-quark, the gluon and the photons, respectively.
The kinematics is fully described in terms of five phase-space
variables $x_1$, $x_2$, $x_3$, $x_4$ and $x_5$ as given explicitly
in eqs. (3.6), (3.9) and (3.10) in ref. \cite{Asatrian:2006ph}.
By identifying $k_1$, $k_2$, $k_3$ and $k_4$ given there with the four-momenta
of the two photons, the $s$-quark and the gluon, respectively, we
easily derive from the information in \cite{Asatrian:2006ph} the
formula for the triple differential decay width. We remind the reader that in this
paper we consider only the range in $s_1$ and $s_2$ with
\[
0 \le s_1 \le 1 \, ; \quad 0 \le s_2 \le 1-s_1 \, ,
\]
which is also accessible to the three-body decay $b \to s \gamma \gamma$.
For this case we obtain
\begin{eqnarray}
\frac{d\Gamma_{1 \to 4}}{ds_1 \, ds_2 \,ds_3} = && 
\frac{(4\pi)^{-6+3\epsilon} \, 2^{-4 \epsilon} \,
  \Gamma[1-\epsilon]}{(1-2\epsilon) \, \Gamma^2[1-2\epsilon]} \,
m_b^{3-6\epsilon} \, s_3^{-\epsilon} \, (s_1 s_2 - s_3)^{-\epsilon} \, 
(1-s_1-s_2+s_3)^{-\epsilon} \times \nonumber \\
&& \int dx_4 \, dx_5 \, [x_4 \, (1-x_4)]^{-\epsilon} \, [x_5 \,
  (1-x_5)]^{-1/2-\epsilon} \, |M_4|^2 \, , 
\label{phasespace4}
\end{eqnarray}
where $x_1$, $x_2$ and $x_3$ (appearing in $|M_4|^2$) are understood to be expressed in terms
of $s_1$, $s_2$ and $s_3$ according to
\be 
x_1=s_1 \, ; \quad x_2 = \frac{s_3}{s_1} \, ; \quad x_3=\frac{s_1s_2-s_3}{(1-s_1)
    \, (s_1-s_3)} \, .
\ee
$x_4$ and $x_5$ vary between zero and one, while $s_3 \in [0,s_1 \, s_2]$. 

\section{Renormalization constants}\label{append:renomalizationconstants}
In this appendix, we collect the  explicit expressions of the renormalization constants needed for the
ultraviolet renormalization in our calculation (see section \ref{sec:virtual}).

\noindent The operator ${\cal O}_{7}$, as well as the $b$-quark mass
contained in this operator are renormalized in the $\MS$ scheme \cite{Misiak:1994zw}:
\be
 Z_{77}^{\MS} = 1 + \frac{4\,C_F}{\epsilon}\frac{\alpha_s(\mu)}{4\pi}
 + O(\alpha_s^2) \quad ; \quad
 Z_{m_b}^{\MS} = 1 - \frac{3\,C_F}{\epsilon}\frac{\alpha_s(\mu)}{4\pi}
 + O(\alpha_s^2)\, .
\ee

\noindent All the remaining fields and parameters are 
renormalized in the on-shell scheme. The on-shell renormalization constant for
the $b$-quark mass is given by
\be
 Z_{m_b}^{\rm OS} =
 1-C_F\,\Gamma(\epsilon)\,e^{\gamma\epsilon}\,
 \frac{3-2\epsilon}{1-2\epsilon}
 \left(\frac{\mu}{m_b}\right)^{2\epsilon}\frac{\alpha_s(\mu)}{4\pi} +
 O(\alpha_s^2)\, .
\ee
while the renormalization constants for the  $s$- and 
$b$-quark fields are
\bea
  Z_{2s}^{\rm OS} &=& 1 + O(\alpha_s^2)\, , \nonumber \\
  Z_{2b}^{\rm OS} &=& 1 - C_F\,\Gamma(\epsilon)\,e^{\gamma\epsilon}\,\frac{3-2\epsilon}{1-2\epsilon}
  \left(\frac{\mu}{m_b}\right)^{2\epsilon}\frac{\alpha_s(\mu)}{4\pi} +
  O(\alpha_s^2)\, .
\eea
The various quantities $\delta Z$ appearing in section
\ref{sec:virtual} are defined to be $\delta Z = Z-1$.


\begin{thebibliography}{99}

\bibitem{Misiak:2006zs}
  M.~Misiak {\it et al.},
  Phys.\ Rev.\ Lett.\  {\bf 98 } (2007)  022002,
  [hep-ph/0609232].

\bibitem{Hurth:2010tk}
  T.~Hurth, M.~Nakao,
  Ann.\ Rev.\ Nucl.\ Part.\ Sci.\  {\bf 60 } (2010)  645,
  [arXiv:1005.1224 [hep-ph]].

\bibitem{Buras:2011we}
  A.~J.~Buras,
  arXiv:1102.5650 [hep-ph].

\bibitem{Simma:1990nr}
  H.~Simma, D.~Wyler,
  Nucl.\ Phys.\  {\bf B344 } (1990)  283.
 

\bibitem{Reina:1996up}
  L.~Reina, G.~Ricciardi, A.~Soni,
  Phys.\ Lett.\  {\bf B396 } (1997)  231,
  [hep-ph/9612387].


\bibitem{Reina:1997my}
  L.~Reina, G.~Ricciardi, A.~Soni,
  Phys.\ Rev.\  {\bf D56 } (1997)  5805,
  [hep-ph/9706253].

\bibitem{Cao:2001uj}
  J.~J.~Cao, Z.J.~Xiao, G.~R.~Lu,
  Phys.\ Rev.\  {\bf D64 } (2001)  014012,
  [hep-ph/0103154].

\bibitem{Chetyrkin:1996vx}
  K.~G.~Chetyrkin, M.~Misiak and M.~Munz,
  Phys.\ Lett.\  B {\bf 400} (1997) 206
  [Erratum-ibid.\  B {\bf 425} (1998) 414],
  [arXiv:hep-ph/9612313].

\bibitem{Melnikov:2005bx}
K.~Melnikov and A.~Mitov, Phys.\ Lett.\ B {\bf 620} (2005) 69,
  [arXiv:hep-ph/0505097].

\bibitem{Asatrian:2006sm}
  H.~M.~Asatrian, T.~Ewerth, A.~Ferroglia, P.~Gambino, C.~Greub,
  Nucl.\ Phys.\  {\bf B762 } (2007)  212,
  [hep-ph/0607316].

\bibitem{Kapustin:1995fk}
  A.~Kapustin, Z.~Lint and H.~D.~Politzer,
  Phys.\ Lett.\  B {\bf 357} (1995) 653,
  [arXiv:hep-ph/9507248].

\bibitem{Gemintern:2004bw}
  A.~Gemintern, S.~Bar-Shalom, G.~Eilam,
  Phys.\ Rev.\  {\bf D70 } (2004)  035008,
  [hep-ph/0404152].

\bibitem{Chang:1997fs}
  C.~-H.~V.~Chang, G.~-L.~Lin, Y.~-P.~Yao,
  Phys.\ Lett.\  {\bf B415 } (1997)  395,
  [hep-ph/9705345].

\bibitem{Hiller:1997ie}
  G.~Hiller, E.~O.~Iltan,
  Phys.\ Lett.\  {\bf B409 } (1997)  425,
  [hep-ph/9704385].

\bibitem{Bosch:2002bv}
  S.~W.~Bosch, G.~Buchalla,
  JHEP {\bf 0208 } (2002)  054,
  [hep-ph/0208202].


\bibitem{Bosch:2002bw}
  S.~W.~Bosch,
  hep-ph/0208203.

\bibitem{Hiller:2004wc}
  G.~Hiller, A.~S.~Safir,
  JHEP {\bf 0502 } (2005)  011,
  [hep-ph/0411344].


\bibitem{Hiller:2005ga}
  G.~Hiller, A.~S.~Safir,
  PoS {\bf HEP2005 } (2006)  277,
  [hep-ph/0511316].

\bibitem{Lin:1989vj}
  G.~L.~Lin, J.~Liu and Y.~P.~Yao,
  Phys.\ Rev.\ Lett.\  {\bf 64} (1990) 1498.

\bibitem{Herrlich:1991bq}
  S.~Herrlich and J.~Kalinowski,
  Nucl.\ Phys.\  B {\bf 381} (1992) 501.

\bibitem{Choudhury:2002yu}
  S.~R.~Choudhury, G.~C.~Joshi, N.~Mahajan, B.~H.~J.~McKellar,
  Phys.\ Rev.\  {\bf D67 } (2003)  074016,
  [hep-ph/0210160].

\bibitem{Aliev:1997uz}
  T.~M.~Aliev, G.~Hiller, E.~O.~Iltan,
  Nucl.\ Phys.\  {\bf B515 } (1998)  321,
  [hep-ph/9708382].

\bibitem{Bertolini:1998hp}
  S.~Bertolini, J.~Matias,
  Phys.\ Rev.\  {\bf D57 } (1998)  4197,
  [hep-ph/9709330].

\bibitem{Bigi:2006vc}
  I.~I.~Bigi, G.~Chiladze, G.~Devidze, C.~Hanhart, A.~Liparteliani, U.~-G.~Meissner,
  hep-ph/0603160.

\bibitem{Devidze:1998hy}
  G.~G.~Devidze, G.~R.~Jibuti,
  hep-ph/9810345.

\bibitem{Aliev:1993ea}
  T.~M.~Aliev, G.~Turan,
  Phys.\ Rev.\  {\bf D48 } (1993)  1176.
  
\bibitem{Xiao:2003jn}
  Z.~J.~Xiao, C.~D.~Lu, W.~J.~Huo,
  Phys.\ Rev.\  {\bf D67 } (2003)  094021, [hep-ph/0301221].

\bibitem{XiuMei:2011iv}
  Q.~XiuMei, W.~Huo, X.~Yang,
  Chin.\ Phys.\  {\bf C33 } (2009)  252,
  [arXiv:1101.2437 [hep-ph]].

\bibitem{Huo:2003cj}
  W.~J.~Huo, C.~D.~Lu, Z.~J.~Xiao,
  hep-ph/0302177.

\bibitem{Chen:2011te}
  H.~Chen, W.~Huo,
  arXiv:1101.4660 [hep-ph].

\bibitem{Ignatiev:2003qm}
  A.~Y.~Ignatiev, G.~C.~Joshi, B.~H.~J.~McKellar,
  Int.\ J.\ Mod.\ Phys.\  {\bf A20 } (2005)  4079,
  [hep-ph/0308126].

\bibitem{Anastasiou:2002qz}
  C.~Anastasiou, L.~J.~Dixon and K.~Melnikov,
  Nucl.\ Phys.\ Proc.\ Suppl.\  {\bf 116} (2003) 193,
  [arXiv:hep-ph/0211141].

\bibitem{Anastasiou:2002yz}
C.~Anastasiou and K.~Melnikov, Nucl.\ Phys.\ B {\bf 646} (2002) 220,
  [arXiv:hep-ph/0207004].

\bibitem{Anastasiou:2003ds}
C.~Anastasiou, L.~J.~Dixon, K.~Melnikov and F.~Petriello, Phys.\ Rev.\ D {\bf
  69} (2004) 094008, [arXiv:hep-ph/0312266].

\bibitem{Laporta:2001dd}
S.~Laporta, Int.\ J.\ Mod.\ Phys.\ A {\bf 15}, 5087 (2000),
  [arXiv:hep-ph/0102033].

\bibitem{Tkachov:1981wb}
F.~V.~Tkachov, Phys.\ Lett.\ B {\bf 100}, 65 (1981).

\bibitem{Chetyrkin:1981qh}
K.~G.~Chetyrkin and F.~V.~Tkachov, Nucl.\ Phys.\ B {\bf 192}, 159 (1981).

\bibitem{Anastasiou:2004vj}
C.~Anastasiou and A.~Lazopoulos, JHEP {\bf 0407}, (2004) 046,
  [arXiv:hep-ph/0404258].

\bibitem{Smirnov:2008iw}
  A.~V.~Smirnov,
  JHEP {\bf 0810 } (2008)  107,
  [arXiv:0807.3243 [hep-ph]].

\bibitem{Remiddi:1997ny}
E.~Remiddi, Nuovo Cim.\  A {\bf 110}, (1997) 1435, [arXiv:hep-th/9711188].

\bibitem{Argeri}
  M.~Argeri and P.~Mastrolia,
  Int.\ J.\ Mod.\ Phys.\  A {\bf 22} (2007) 4375,
  [arXiv:0707.4037].

\bibitem{Asatrian:2006ph}
  H.~M.~Asatrian, A.~Hovhannisyan, V.~Poghosyan, T.~Ewerth, C.~Greub and T.~Hurth,
  Nucl.\ Phys.\  B {\bf 749} (2006) 325,
  [arXiv:hep-ph/0605009].

\bibitem{Misiak:1994zw}
M.~Misiak and M.~Munz, Phys.\ Lett.\ B {\bf 344} (1995) 308,
  [arXiv:hep-ph/9409454].



\end{thebibliography}
\end{document}